# Threat Actor Type Inference and Characterization within Cyber Threat Intelligence

**Vasileios Mavroeidis**
University of Oslo
Oslo, Norway
vasileim@ifi.uio.no

**Ryan Hohimer**
DarkLight Inc.
Richland, Washington, United States
ryan.hohimer@darklight.ai

**Tim Casey**
Intel Corp.
Chandler, Arizona, United States
tim.casey@intel.com

**Audun Jøsang**
University of Oslo
Oslo, Norway
audun.josang@mn.uio.no

**Abstract:** As the cyber threat landscape is constantly becoming increasingly complex and polymorphic, the more critical it becomes to understand the enemy and its modus operandi for anticipatory threat reduction. Even though the cyber security community has developed a certain maturity in describing and sharing technical indicators for informing defense components, we still struggle with non-uniform, unstructured, and ambiguous higher-level information, such as the threat actor context, thereby limiting our ability to correlate with different sources to derive more contextual, accurate, and relevant intelligence. We see the need to overcome this limitation in order to increase our ability to produce and better operationalize cyber threat intelligence. Our research demonstrates how commonly agreed-upon controlled vocabularies for characterizing threat actors and their operations can be used to enrich cyber threat intelligence and infer new information at a higher contextual level that is explicable and queryable. In particular, we present an ontological approach to automatically inferring the types of threat actors based on their personas, understanding their nature, and capturing polymorphism and changes in their behavior and characteristics over time. Such an approach not only enables interoperability by providing a structured way and means for sharing highly contextual cyber threat intelligence but also derives new information at This research work was supported by the research project CyberHunt (Grant No. 303585) funded by the Research Council of Norway.

machine speed and minimizes cognitive biases that manual classification approaches entail.

## 1. INTRODUCTION

Cyber threat intelligence (CTI) is undeniably an essential element for building a robust security posture against adversarial attacks. Establishing a threat intelligence program allows security teams to benefit from increased situational awareness, and thus minimize their organizations' attack surfaces. Evidence-based knowledge of both adversary dynamics and an organization's attack surface can support anticipatory threat reduction. Organizations follow a process of increasing maturity with respect to their cyber capability, transitioning from manual and reactive approaches to more automated and proactive.

Proactive cyber defense is intelligence-driven and focuses on providing awareness and preparing an organization against anticipated attacks. Every adversarial attack can be decomposed into elements that provide information about the who, what, where, when, why, and how. The *who*, commonly known as attribution, identifies the individual, group, organization, or nation that conducted the adversarial operation. The *what* reflects the scope of the attack. The *where* relates to the attack's direction, such as where it is coming from and its target – an organization, industry, or country. The *when* can be perceived as the timestamp of the attack and can be deterministic or probabilistic. The *why* is equivalent to motivation and designates the goals and the objectives of the adversary. The *how* is made up of the tactics, techniques, and procedures (TTPs) employed by the adversary for conducting the operation. Collectively, these factors provide insight into how adversaries plan, conduct, and sustain their operations.

Attribution is typically a challenging task requiring direct evidence through principled and systematic analysis which correlates multiple internal and external data sources and threat intelligence. Such a process identifies and maps TTPs and associated tools and infrastructure to known sources of similar attacks. However, threat actors intend to remain unidentified and employ deception and obfuscation techniques that can lead to incorrect attribution or weakening the possibility of correctly associating a particular activity with a known adversary. For example, the Russia-backed group Turla (also known as Waterbug) was discovered to be using the infrastructure and malware of APT34 (also known as OilRig), an Iranian threat group [1]. Nevertheless, many times, a threat actor profile is created and linked to one or more adversarial operations based on common identifiable properties without actual attribution, meaning that the adversary's real-world identity remains unknown.

Capturing high-level information, such as the motives behind an adversarial operation and contextualizing technical findings; for example, by estimating the sophistication level, skills, and resources needed to plan and execute the attack, can characterize the perpetrator and infer its nature even when direct attribution has not been achieved. The opposite is also plausible. The nature of a perpetrator reflects its capability, persistence, and motives. In addition, in a threat landscape that has become very diversified and hybridized, the importance of portraying adversaries and their nature as threat actor types is apparent. Threat actors are continuously evolving and are becoming polymorphic with multiple motivations and goals. Existing approaches in characterizing threat actors and their operations mostly fall under the category of regular intelligence reports that fail to capture information in a specific representation format that both humans and machines can interpret. On the other side, lies purely technical information intended to be consumed directly by cyber defense products.

A wide range of threat actor types exists, ranging from disgruntled employees to organized cyber crime and nation-state-backed groups. Threat actors have specific traits common to most of their behaviors. For example, an employee with a grudge against their organization is motivated by disgruntlement. In contrast, a state-sponsored group may aim to achieve dominance over another nation for geopolitical reasons. To operationalize this type of characterization, we need to satisfy two criteria. First, the definitions of actor types must be unambiguous, and second, we must characterize them using a set of attributes that enables robust, reliable enumeration and inference.

This research reflects the operational and strategic benefits derived from semantically portraying threat actors as threat actor types (e.g., nation-state, hacktivist, terrorist, organized cyber crime) to understand the actors' nature and capture polymorphism and changes in their behavior and characteristics over time. Furthermore, we present an ontological system for threat actor type inference which relies on a standard set of attributes for characterizing threat actors and their operations. Axioms (expressions) capture domain knowledge regarding the composition of threat actor types based on their defining attributes. The presented approach can augment existing static enumerative approaches for threat actor type classification with a flexible generative system based on the logic encapsulated in the ontologies. Such an approach enables machine understanding and logical reasoning based on that understanding with transparent and explicable results. The proof-of-concept ontology we engineered utilizes Casey's Threat Agent Library (TAL) [2]. The original TAL typology has been refined and can be updated further to reflect a more contemporary description of threat actor types and their defining attributes.

A semantically expressed threat actor typology based on a set of standard characterization attributes provides the following advantages.

- Based on commonly agreed-upon definitions, a machine-understandable interpretation of threat actor types and their defining attributes eliminates ambiguity regarding their meaning by annotating their unique characteristics. The term *commonly* above refers to the need for interoperability. A standard vocabulary and representation for threat actor types can be integrated across different technologies, such as threat intelligence platforms and threat intelligence sharing languages, and used when generating threat reports. For example, people often interpret seemingly simple terms such as *hacktivist* differently. Correlating a threat actor type with an operation is then subject to fallacies when the semantics for what comprises a particular type are not in place. This makes shareable information inaccurate and contradictory since different entities may have different interpretations of the same term leading to inconsistent threat actor profiles.

- Representing domain knowledge in a declarative form, such as axioms and facts, can enable automatic inference via the ability of machines to reach a conclusion based on evidence. In this research, axioms capture the unique attribute combinations that characterize different threat actor types. Using a description logics reasoner, also known as an inference engine, instances of threat actors can be programmatically examined to infer their type. Automatic inference also speeds up traditional analytical processes that require competing hypotheses about the adversary's type to be tested.

- Polymorphism and changes in threat actor behavior over time are becoming common, with adversaries being influenced by different motivations and goals. Some threat actors evolve in nature and gradually engage in larger-scale and more complex operations. In contrast, others pause their operations, disappear, or even go through organizational changes like establishing new units. It is essential for the threat intelligence community to recognize and formally represent polymorphism and behavioral changes over time so that threat actor profiles can evidentially account for more than one threat actor type (Figure 1). For example, as presented in Section 5, the state-sponsored Lazarus Group has engaged in activities not only motivated by geopolitical reasons to achieve dominance over other nations by conducting stealthy cyber-espionage campaigns but also for nationalistic reasons and revenge by engaging in destructive hacking, as well as for financially motivated reasons by conducting bank heists possibly to fund their operations. As discussed later, available threat actor knowledge bases appear to fail to capture polymorphism and behavioral changes, resulting in monolithic representations that lack evidence-based relationships concerning the derivation of their characterization. In addition, most of the time, the characterizations are based on proprietary works that are also ambiguous due to nonexistent or insufficient definitions. Ambiguity and imprecision create confusion and diminish the value of intelligence in cyber operations.

**FIGURE 1:** SEMANTIC MODELING OF THREAT ACTOR POLYMORPHISM

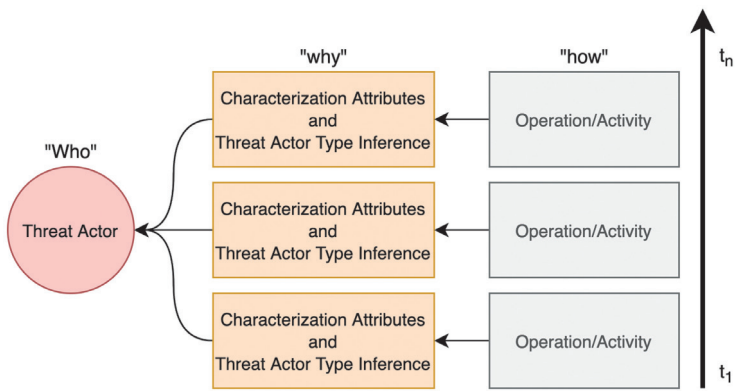

- The definition and utilization of characterization attributes (e.g., motivations, goals, objectives, visibility) can contextually enrich cyber threat intelligence and enable granular querying of higher contextual precision to answer complex questions. The derived intelligence can provide defenders with increased situational awareness and thus allow them to better prioritize their defense efforts according to their most relevant threats.

The rest of the paper is organized as follows. Section 2 introduces the Threat Agent Library [2] that was referenced to create a prototype ontology for threat actor type inference, and presents and analyzes different threat actor knowledge bases with respect to how they handle high-level contextual information in terms of ambiguity, structured shareability, explainability, and most importantly operationalization ability. Additionally, Section 2 discusses how the Structured Threat Information eXpression (STIX) language deals with interpreting threat actor polymorphism. Section 3 discusses knowledge representation and ontology engineering within the cyber threat intelligence domain, and annotates how ontology inference can provide defenders with additional information and insights at machine speed. Section 4 presents an ontology for threat actor characterization and threat actor type inference. Section 5 validates the proposed concept's efficacy and presents a use-case analysis where the ontology presented in Section 4 is used to infer threat actor types automatically. Furthermore, Section 5 demonstrates the potential of characterization attributes in providing highly contextual and queryable cyber threat intelligence. Finally, Section 6 concludes the paper.

## 2. BACKGROUND INFORMATION

### A. Threat Agent Library

Introduced in 2007, the Threat Agent Library (TAL) [2] is a set of definitions and descriptions to represent significant threat agent categories, or as termed in this paper, threat actor types. The TAL was developed to support risk management processes by simplifying the identification of threat agent archetypes that pose the most significant risk to specific assets (Figure 2). Based on the available information on each archetype class, an organization can get an insight into current adversarial activities and consequently take action to improve its security posture. The library (Table I) enumerates twenty-one archetypes (e.g., government spy, radical activist, untrained employee, disgruntled employee) and their associated defining attributes: access, outcome, limits, resources, skills, objective, visibility, and motivation. The defining attributes reflect the typical characteristics of each threat actor type.

**FIGURE 2:** RISK ASSESSMENT USING THE THREAT AGENT LIBRARY

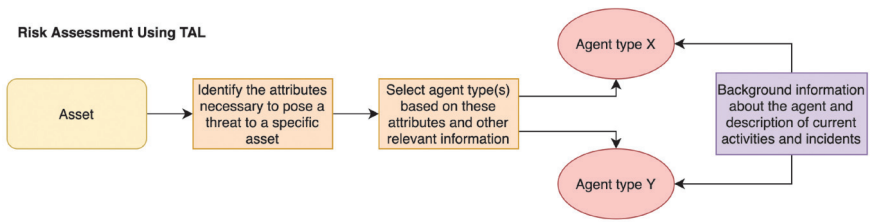

This research presents a proof-of-concept ontological representation of TAL, with minor improvements, for automatically inferring threat actor types from cyber threat intelligence instances (objects). The decision to use TAL is based on its assessment of combinations of characterization attributes that uniquely identify different threat actor types. Further, we emphasize the importance of having a set of standard characterization attributes to contextualize cyber threat intelligence, thereby making it more actionable and relevant. We also argue that modeling approaches should be temporal-based to capture threat actor polymorphism and behavioral changes over time. As presented in the next sections, available threat actor knowledge bases struggle to capture such formalisms resulting in contextual loss and ambiguity.

## B. Threat Actor Characterization Using STIX 2.1

Structured Threat Information eXpression (STIX) is a schema that defines a taxonomy for cyber threat intelligence. We discuss and analyze STIX version 2.1 [3] for two reasons. First, because of its ability to describe threat actors, threat actor activity, and their associated characteristics in a machine-readable format, and second, because it has been embraced as the standard representation format for sharing cyber threat intelligence in a structured manner.

The **STIX Threat Actor object** aggregates information about threat actors, such as their goals, motivations, sophistication, resource-level, and type. Additionally, it utilizes relationship objects to reference objects that represent the actual identity behind a threat actor (be it a human or organization), the tools that the actor has been known to use or used in a specific attack, the patterns of attack that the actor is known to follow, the location where the actor is believed to be, infrastructure both owned and compromised that the actor is known to use, as well as attributes about the actor that help characterize them. This is an object of high value in proactive cyber defense where strategic, operational, and tactical cyber threat intelligence play a significant role. Figure 3 presents the STIX threat actor object with its characterization attributes and relationships with other objects.

**FIGURE 3:** STIX THREAT ACTOR OBJECT

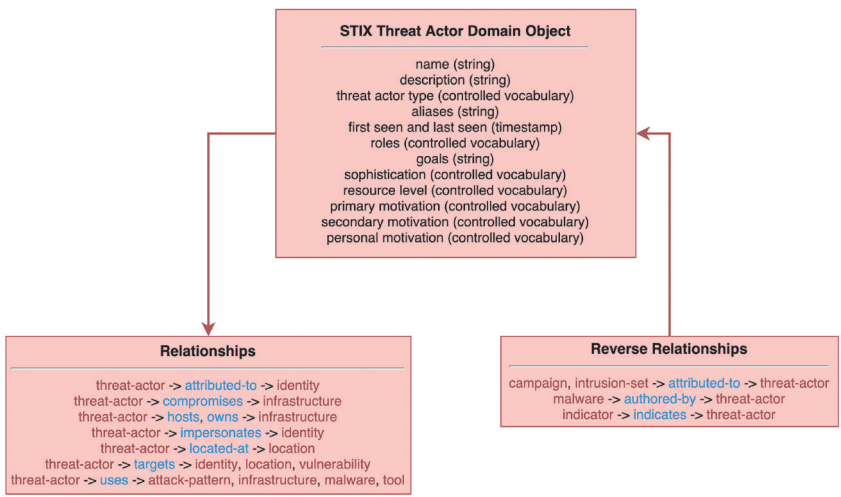

A critical aspect that the STIX threat actor object does not account for is capturing and semantically representing behavioral polymorphism in a temporal manner, as in the case where a threat actor is conducting different operations than what is known, reflecting a possible change to its primary or secondary motivations and goals. Furthermore, the characterization attributes of the threat actor object do not hold any direct relationships with other objects to justify the existing characterization. This is especially the case when a threat actor object has more than one value populated for an attribute (e.g., a threat actor that accounts for more than one threat actor type). Also, some of the STIX vocabularies used for characterizing adversaries are ambiguous because they lack definitions. The generation of the threat actor type attribute is a manual and subjective process prone to human fallacies. For example, a threat actor object with the populated threat actor type value *nation-state* and resource-level *individual* (limited resources) is unlikely to be correct but is deemed a valid STIX statement. This reflects the advantage of utilizing an automated generative threat actor type inference approach (Section 4) for augmenting existing manual approaches.

## C. Threat Actor Knowledge Bases

A knowledge base is a collection of information about a particular subject area that can be used to support decision-making and draw conclusions. A knowledge base with information about threat actors' capabilities, goals and motivations, and past and ongoing activities can inform prevention and response strategies. Unstructured knowledge bases can be a simple aggregating system such as a collection of threat reports. At a basic level, the development of a structured knowledge base requires a schema that defines its structural composition, information sources for populating the knowledge base, and optimally controlled vocabularies for additional context and granular searchability. Describing a threat actor with high-confidence demands processing, correlating, analyzing, and integrating different relevant intelligence sources.

This section presents a set of open-source threat actor knowledge bases, and analyzes their structural composition with respect to how easy it is to operationalize them in the context of finding information relevant to our needs.

**MITRE ATT&CK** [4] is a knowledge base of known adversary tactics and techniques based on openly available analyzed activity. It is a valuable resource to better understand observed adversarial behavior, and it can be used for multiple purposes, such as for adversary emulation, behavioral analytics, cyber threat intelligence enrichment, defense gap assessment, red teaming, and SOC maturity assessment [5]. ATT&CK matrices exist about adversary behavior targeting enterprise environments, mobile, and industrial control systems. Moreover, information pertinent to the software adversaries use, mitigation techniques, procedure examples, and detection recommendations are also available. Further, the associated PRE-ATT&CK matrix focuses on operational techniques known to be utilized before an attacker exploits a particular target network or system.

Of particular importance is the available ATT&CK Groups knowledge base, a list of known adversaries and their associated techniques and software tools. Figure 4 shows the main components of ATT&CK and their relationships.

**FIGURE 4:** ATT&CK MODEL RELATIONSHIPS – REDESIGNED FROM [5]

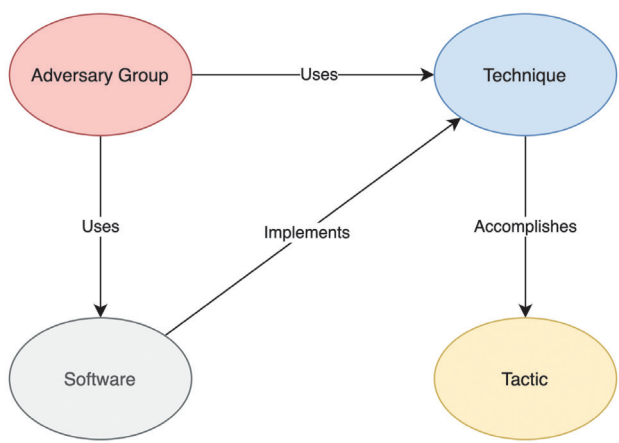

One way of getting started with ATT&CK is identifying adversarial groups relevant to an organization, based on whom they have previously targeted, such as similar organizations within the same sector, and then look at their TTPs [6]. TTPs that are commonly used can be prioritized for detection and mitigation. However, the ATT&CK Groups knowledge base lacks proper structurality and relationships between adversaries and their targets and between adversaries and their motivations. Information such as targeted countries and sectors and threat group motivations is embedded within the general description of a group and can be unstructurally searched using the ATT&CK portal. However, the vocabularies utilized to specify a group's targets and their motivations are not available, limiting searchability, and consequently, the ability to extract more relevant information. Synergistically, structuring the available information, establishing relationships between concepts, and utilizing a set of standard characterization attributes and other common vocabularies can facilitate more informed and targeted queries over the knowledge base, resulting in getting more relevant, and maybe otherwise missed TTPs to prioritize.

The description of APT19[1] is a good example of unstructured populated information regarding industries the group has targeted.

> APT19 is a Chinese-based threat group that has targeted a variety of industries, including defense, finance, energy, pharmaceutical, telecommunications, high tech, education, manufacturing, and legal services. In 2017, a phishing campaign was used to target seven law and investment firms.

Similarly, the description of APT38[2] is a good example of unstructured populated information regarding a group's motivations.

> APT38 is a financially motivated threat group that is backed by the North Korean regime. The group mainly targets banks and financial institutions and has targeted more than 16 organizations in at least 13 countries since at least 2014.

The **Threat Actor Encyclopedia** [7] is an effort from Thailand's Computer Emergency Response Team (ThaiCERT) to create a knowledge base of threat group profiles by aggregating, processing, and structuring open-source intelligence. As in other efforts, we observed ambiguity and confusion regarding the interpretation and use of characterization attributes. For instance, the threat actor encyclopedia's motivation vocabulary includes the terms *information theft and espionage*, *financial crime*, *financial gain*, and *sabotage and destruction*. Definitions of the above terms have not been provided, making it difficult, for example, to understand the contextual difference between financial gain and financial crime. It can also be argued that *information theft and espionage*, *sabotage and destruction*, and *financial crime* are not motivation types but operation types or intended effects.

The Malware Information Sharing Platform (MISP) is an open-source threat intelligence platform for collecting, storing, and sharing information about cyber security incidents [8]. Due to its open-source nature and modular architecture, the platform can integrate intelligence clusters that, in many cases, are community-driven efforts and can be used to enrich events and attributes. The **MISP Threat Actor cluster**[3] is a knowledge base of threat groups. The cluster's structural composition is an array of threat group objects that capture information related to the groups, such as name and related aliases, a description, targeted countries and sectors (e.g., private, military, government), their affiliated countries and sponsors, attribution confidence, incident types (e.g., espionage, sabotage, or defacement), references relating to the captured knowledge, relations with other groups and operations, and associated malware. A subset of the elements has been derived from the Council on Foreign

---

1 https://attack.mitre.org/groups/G0073/
2 https://attack.mitre.org/groups/G0082/
3 https://github.com/MISP/misp-galaxy/blob/main/clusters/threat-actor.json

Relations Cyber Operations[4] vocabulary used for reporting cyber incidents. Like the rest of the knowledge bases investigated, the MISP Threat Actor cluster could benefit from introducing a more expressive structured representation. Currently, multiple characterization attributes are included only in the general description of a threat actor object, making it difficult to parse the information via automated means. For instance, in the example below, the description captures information regarding the motivations, objectives, targeted countries, and the types of operations a group has been observed conducting.

> Libyan Scorpions is a malware operation in use since September 2015 and operated by a politically motivated group whose main objective is intelligence gathering, spying on influential and political figures, and operating an espionage campaign within Libya.

Moreover, the use of different non-standardized vocabularies for enriching the knowledge base and the integration of different intelligence sources for providing additional context introduces ambiguity and confusion. The two shortened examples presented below indicate the importance of utilizing a set of standard characterization attributes with accurate definitions and vocabularies for optimally resolving ambiguity and operationalizing the provided intelligence.

In the example below, *espionage* is used both to describe an incident type and a motive. Additionally, definitions for the available terms are not in place, increasing the probability of misusing the vocabularies.

```
{
   "description": "Anchor Panda is an adversary that CrowdStrike has tracked
extensively over the last year targeting both civilian and military maritime
operations...",
   "meta": {
     "attribution-confidence": "50",
     "cfr-suspected-state-sponsor": "China",
     "cfr-suspected-victims": ["United States", "..."],
     "cfr-target-category": ["Government", "..."],
     "cfr-type-of-incident": "Espionage",
     "country": "CN",
     "motive": "Espionage",
     "refs": ["..."],
     "synonyms": ["APT14"]
   },
   "value": "Anchor Panda"
}
```

---

4   https://www.cfr.org/cyber-operations/

337

In the example below, the motive of the group is defined as Hacktivists-Nationalists, which is reminiscent of a threat actor/group type rather than a motive that influences the actions of an actor.

```
{
   "description": "Turkish nationalist hacktivist group that has been active for
roughly one year...The group carries out distributed denial-of-service (DDoS)
attacks and defacements against the sites of news organizations and governments
perceived to be critical of Turkey's policies or leadership, and purports to act in
defense of Islam",
   "meta": {
     "attribution-confidence": "50",
     "country": "TR",
     "motive": "Hacktivists-Nationalists",
     "synonyms": ["Lion Soldiers Team", "..."]
   },
   "value": "Aslan Neferler Tim"
}
```

## 3. KNOWLEDGE REPRESENTATION AND ONTOLOGY

Knowledge representation conceptualizes an understanding of the world. It can provide a view of a particular domain of interest and capture that knowledge in a formal representation so that a computer system can utilize it to solve complex tasks, such as inferring new critical information. An ontology is a formalism of knowledge representation that encodes knowledge about a particular domain. An ontology is machine-understandable, holds formal semantics that carry meaning, and allows for reasoning. Formal semantics and logic ensure that the meaning of a concept is unambiguous. An ontology is defined using a knowledge representation language, such as the Web Ontology Language (OWL). An OWL ontology consists of the following three syntactic categories [9]: a sequence of logical *axioms* (statements) that are asserted to be true in the domain being described, *expressions* that represent complex notions in the domain being described (e.g., a class expression describes a set of individuals in terms of the restrictions on the individuals' characteristics), and *entities* such as classes, properties, and individuals, that constitute the basic elements of an ontology. A class represents a concept and provides the means for grouping resources with similar characteristics. For instance, a *threat actor* class can group all known adversaries. Subclasses represent concepts that are more specific than a superclass. For instance, the class *threat actor* can decompose into subclasses that capture a threat actor's intent, such as *hostile* or *nonhostile*, and again decompose into subclasses that define hostile or nonhostile types, such as *nation-state*, *civil activist*, and *untrained employee*. Taking the Lazarus Group as an example and based on available information, it can be classified as a *nation-state* adversary, a subclass of the *hostile* class. The *hostile* class is a subclass of the *threat actor type* class, indicating that the nation-state-backed group Lazarus is an instance of a hostile threat actor. The functional syntax of this example is shown below, with Figure 5 providing an illustration.

```
Declaration ( Class( :ThreatActorType ) )
Declaration ( Class( :Hostile ) )
Declaration ( Class( :NonHostile ) )
Declaration ( Class( :NationState ) )
Declaration ( Class( :UntrainedEmployee ) )
SubClassOf ( :Hostile :ThreatActorType )
SubClassOf ( :NationState :Hostile )
SubClassOf ( :NonHostile :ThreatActorType )
SubClassOf ( :UntrainedEmployee :NonHostile )
Declaration ( NamedIndividual( :LazarusGroup ) )
ClassAssertion ( :NationState :LazarusGroup )
```

**FIGURE 5:** EXAMPLE ILLUSTRATION OF ONTOLOGY CLASSES AND SUBCLASSES

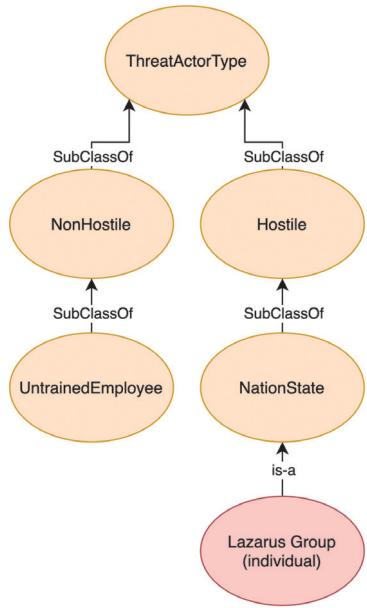

Properties define relationships between individuals (object properties) or between individuals and data type literals (data type properties). For instance, as described in the example provided in Section 2.D, APT38 is a financially motivated threat group that is backed by the North Korean regime. In addition, APT38 is also known as Stardust Chollima by Crowdstrike [10] and as BlueNoroff by Kaspersky [11]. The relation of APT38 with a particular defining motivation and other aliases can be captured by creating relevant object properties and formulating semantic triples. A triple is a set of three entities that codify a statement in the form of subject-predicate-object. This principle is illustrated in Figure 6, where the arcs represent relations (object properties – predicates), and the ellipticals represent individuals.

**FIGURE 6:** SEMANTIC REPRESENTATION OF APT38

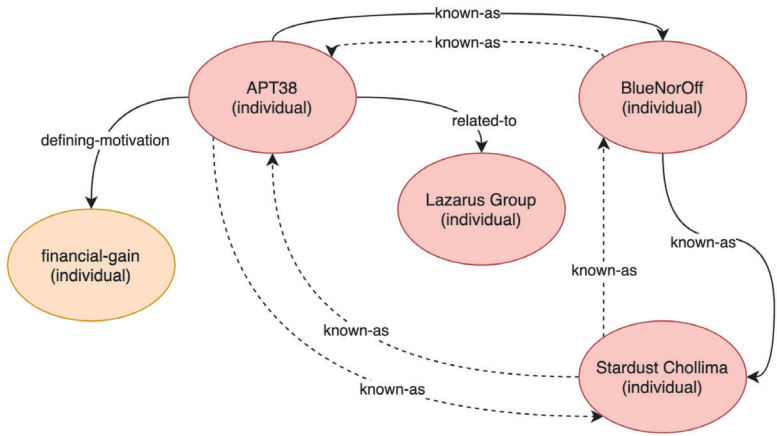

OWL offers expressive constructs for reasoning based on description logics. For example, the defined object property *known-as* is bidirectional when declared symmetric and allows traversing information when declared transitive. Property declarations can compensate for missing arcs in a knowledge base. A reasoner can parse the knowledge base and infer new information. In the example illustrated in Figure 6, the symmetric property *known-as* allows inferring that APT38 is known as BlueNoroff and the opposite, such as that BlueNoroff is known as APT38. Furthermore, because of transitivity, a reasoner infers that StarDust Chollima is also known as APT38 (dashed arc) even though it was not directly defined. Ontological axioms, expressions, and constructs can infer information based on causal relationships. For instance, a reasoner will not infer that a threat actor is of *nation-state* type when the resource-level property is not populated with the value *government*, according to the class expression that encodes what a nation-state threat actor comprises.

# 4. A DOMAIN ONTOLOGY FOR THREAT ACTOR PROFILING

This section presents a domain ontology for threat actor profiling and actor type inference based on the Threat Agent Library (TAL) [2]. TAL defines threat actor type attributes through controlled vocabularies, such as motivation, access, outcome, limits, resources, skills, objectives, and visibility, and when used collectively, these identify the unique characteristics of each threat actor type. Threat actor types refer to categories that adversaries can be classified into, such as spy, civil activist, and nation-state. In TAL, *threat agent* denotes a class of threat actor and is synonymous with *threat actor type*. The definitions of the TAL terms can be found in [2] and [12].

To develop the ontology, we slightly refined TAL to increase its expressiveness and resolve ambiguities that could otherwise affect ontological assertions and inferencing. TAL's threat actor types and their associated defining attributes are shown in Table I. The table's key takeaways are: TAL comprises twenty-one unique threat actor type categories and their associated characteristics based on eight attributes. The motivation attribute was added to the library in later work [12]. The shaded cells in the second column of Table I refer to either minor nonbreaking attribute modifications that resolve ambiguity concerning their ontological use, or attribute updates that allow for more flexible use. For instance, the individualistic motivation Personal Financial Gain has been replaced with Financial Gain to allow more flexible characterization, meaning that the property can now be used to characterize groups and not only individuals, such as organized cyber crime groups that operate mainly for profit, indicating financially motivated actors.

**TABLE I:** THREAT AGENT LIBRARY – REDESIGNED FROM [2]

| | | Non-Hostile | | | Hostile | | | | | | | | | | | | | |
|---|---|---|---|---|---|---|---|---|---|---|---|---|---|---|---|---|---|---|
| | | Reckless Employee | Untrained Employee | Information Partner | Anarchist | Civil Activist | Competitor | Corrupt Government Official | Data Miner | Disgruntled Employee | Government Cyberwarrior | Government Spy | Internal Spy | Irrational Individual | Legal Adversary | Mobster | Radical Activist | Sensationalist | Terrorist | Thief | Vandal | Vendor |

A high-level illustration of the ontology is presented in Figure 7. The threat actor type and characterization attribute classes enumerate possible values using individuals (instances). For example, the *visibility* attribute comprises four individuals that define different levels of visibility: clandestine, covert, opportunistic, and overt.

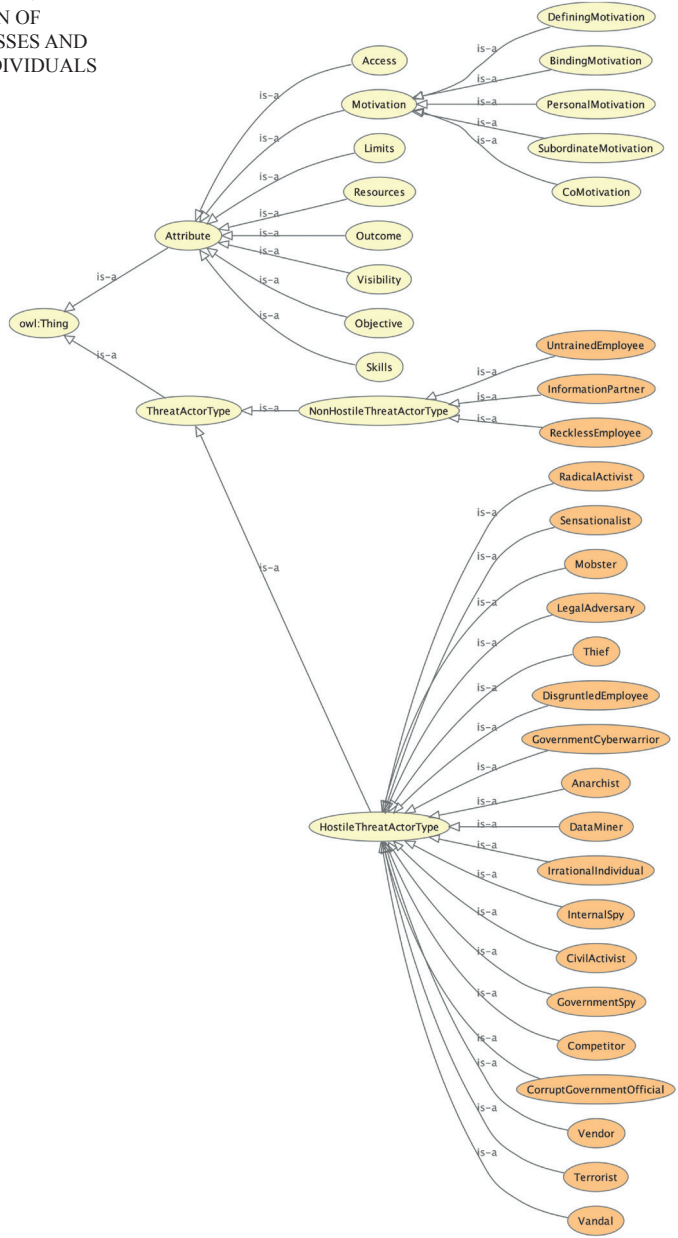

**FIGURE 7:** HIGH-LEVEL REPRESENTATION OF ONTOLOGY CLASSES AND ASSOCIATED INDIVIDUALS

Object properties relate individuals to individuals. For example, an individual (object) that describes an adversarial operation can have a relationship to a motivation that is believed to influence the attack, such as the desire to achieve dominance. This can be expressed using the object property *hasDefiningMotivation*, deriving a semantic triple (*subject-hasDefiningMotivation-dominance*).

In addition, the ontology can automatically infer threat actor types, decreasing the human biases entailed in traditional manual classification and decision-making processes, by capturing the existing domain knowledge within ontology expressions (axioms) that characterize threat actor types based on combinations of the attributes mentioned earlier. An example expression that captures the combination of attributes comprising a nation-state-backed actor (government cyberwarrior based on TAL) is shown below in Manchester syntax.

```
((hasVisibilityAttribute some Visibility) or
(hasVisibilityAttribute value visibility:dontCare))
and ((hasObjectiveAttribute value objective:damage) or
(hasObjectiveAttribute value objective:deny) or
(hasObjectiveAttribute value objective:destroy))
and ((hasOutcomeAttribute value outcome:damage) or
(hasOutcomeAttribute value outcome:embarrassment))
and (hasAccessAttribute value access:external)
and (hasDefiningMotivationAttribute value motivation:dominance)
and (hasLimitsAttribute value limits:extraLegalMajor)
and (hasResourcesAttribute value resources:government)
and (hasSkillsAttribute value skills:adept)
```

Objects with populated attributes that fulfill expression requirements (equivalency) are classified as threat actor types in an automated manner near real-time by a description logics reasoner. As demonstrated in Section 5, polymorphic threat groups can be attributed to more than one threat actor type, compared to traditional enumerative approaches that use mutually exclusive lists and lead to contextual loss. The suggested approach does not prohibit an analyst from manually classifying a threat actor as a specific type or populating other attributes (open world assumption). Changes to the defining characterizations of threat actor types can be reflected by updating the ontology expressions. To enable temporality, the characterization attributes of a threat actor instance are populated using an individual object (instance) that connects with other related instances (e.g., malicious activity or identity) using relationships (Figure 8). Temporality-based knowledge representation can justifiably reflect shifts and polymorphism in adversarial behavior.

**FIGURE 8:** TEMPORALITY-ENHANCED SEMANTIC MODELING OF THREAT ACTOR POLYMORPHISM

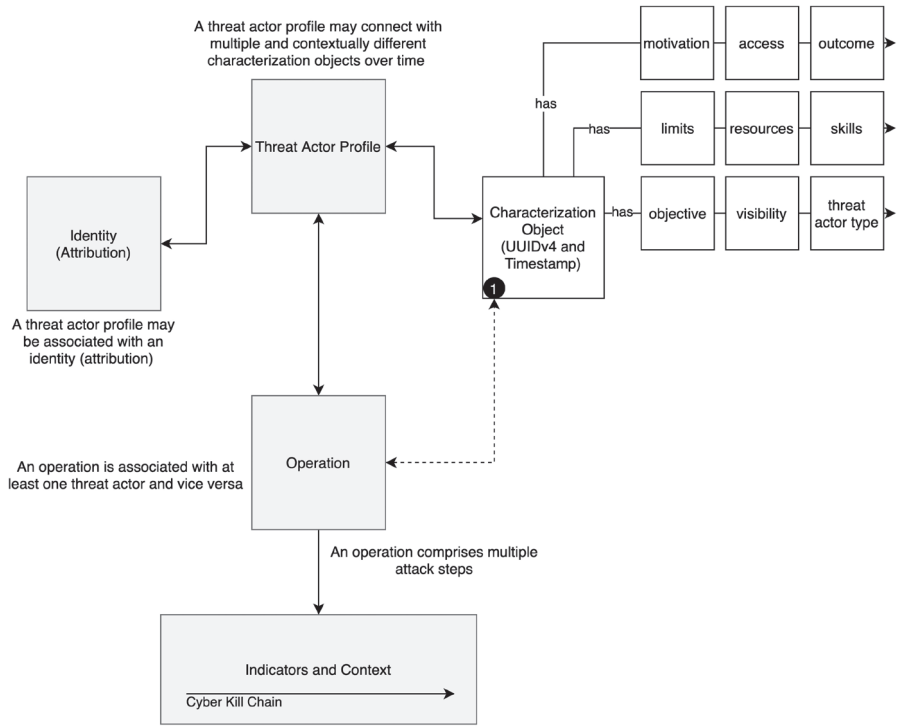

## 5. THE LAZARUS GROUP USE CASE

In this section, we utilize the ontology presented in Section 4 to model the Lazarus Group for the purpose of inferring threat actor types automatically. We demonstrate how a standardized set of characterization attributes for describing adversary capability and behavior makes cyber threat intelligence more contextual and queryable and makes it possible to derive new information at machine speed by utilizing a reasoner. We apply a top-down modeling approach to open-source information about operations believed to have been conducted by the Lazarus Group. Even though an attribution of high confidence has been achieved and the capabilities and sophistication of the Lazarus Group are known, we characterize the operations (use cases) based on their individual characteristics. A top-down modeling approach uses existing knowledge and historical data to create a threat actor profile and is more accurate and contextual than a bottom-up approach, which derives intelligence from early-stage ongoing analyses of cyber attacks. Nevertheless, both modeling methods should follow an evidence-based approach by establishing direct relationships between the characterization attributes and the instances of operations the information has been derived for robust, explicable, and temporal-enabled threat intelligence.

According to the MITRE ATT&CK Groups knowledge base[5]:

> The Lazarus Group is a threat group that has been attributed to the North Korean government. North Korean groups are known to have significant overlap, and the name Lazarus Group is known to encompass a broad range of activity. Some organizations use the name Lazarus group to refer to any activity attributed to North Korea, whereas other organizations track North Korean clusters or groups such as Bluenoroff, APT37, and APT38 separately.

According to the Council on Foreign Relations[6]:

> The Lazarus Group targets and compromises entities primarily in South Korea and South Korean interests for espionage, disruption, and destruction. It has also been known to conduct cyber operations for financial gain, including targeting cryptocurrency exchanges.

The descriptions above are indicative of a polymorphic threat. Based on TAL, an ontological equivalency expression of a nation-state threat actor (government cyberwarrior) identifies the following characteristics:

- *access → external*
- *visibility → any-opportunistically*
- *objective → deny-destroy-damage*
- *limits → extra-legal, major*
- *outcome → damage, embarrassment*
- *defining motivation → dominance*
- *skills → adept*
- *resources → government*

Establishing formal threat actor type definitions using a set of machine-readable characterization attributes equips defenders with a queryable representation that can derive explicable intelligence.

The Lazarus Group is known to have been active for more than a decade and is an example of an adversary that has exhibited polymorphism and increased operational sophistication over time. The nation-state-backed group has engaged in multiple cyber espionage, destructive, disruptive, and financially motivated operations. For example, the DarkSeoul attack on March 20, 2013, targeted South Korean news agencies and banks, causing significant damage to the affected entities by wiping the hard drives of tens of thousands of computers. At an early stage, Symantec stated that the actual motives for the attacks were unclear and added that they might be part of either a clandestine attack or the work of nationalistic hacktivists taking issues into their own hands in response to political tensions on the Korean Peninsula [13]. In a report [14], McAfee, after analysis, remarked that an attack which was initially perceived as an unsophisticated incident of cyber vandalism or hacktivism had actually grown out of a sophisticated multi-year covert cyber espionage campaign that this time was indeed intended to damage, cause disruption, and potentially harvest information. Table I identifies the defining characteristics of a cyber vandal and radical activist according to TAL.

---

5   https://attack.mitre.org/groups/G0032/
6   https://www.cfr.org/cyber-operations/lazarus-group

The threat actors NewRomanic Cyber Army Team and Whois Team, who claimed responsibility for the attacks in South Korea, were later discovered to be a fabrication to mask the real source of the attack. In addition, Marpaung and Lee explained that DarkSeoul was a low-tech threat compared to advanced persistent threats that nation-state groups typically perform [15].

By structuring the information about the DarkSeoul attack, the following characterization attributes emerge. The threat actor was external to the targeted entities (*access → external*) and conducted a large-scale covert operation (*visibility → covert*) which caused destruction, disruption, and possibly harvested information (*objective → destroy, damage*, and maybe *copy*). Based on the attack type and impact, we can conclude that the actor took no account of the law (*limits → extra-legal major*) and that its primary goal was large-scale data destruction with a sequential impact on the affected entities' operations (*outcome → damage*). This type of attack reflects a motivation to achieve dominance over another party, or as in this case, over another nation (*defining motivation → dominance*). Furthermore, what was initially perceived as an unsophisticated attack due to the raw destructive nature of the payload was, in fact, a coordinated strike against multiple entities delivered with precision and planning commonly associated with state-sponsored intrusion campaigns [14] (*skills → adept*), (*resources → government*). Based on the above characterization, a reasoner would infer that a government cyberwarrior conducted the operation, otherwise known as nation-state threat actor type. It is worth noting that the contextual characterization of the DarkSeoul attack in this particular case takes into account information about a set of individual attacks all described in one object, thus indicating a relatively high-level sophistication, which in turn is a factor for estimating the skills and resources required for conducting the attacks. Exemplifying each incident separately would populate objects that a reasoner would infer as the threat actor type (cyber) vandal. The attributes such as motivation, outcome, objectives, and visibility highly overlap between the vandal and government cyberwarrior (nation-state) types. Other attributes such as skills, resources, and limits are dissimilar and annotate the differences in capability between the two types. The attribution of the DarkSeoul attack confirmed that it was planned and executed by a nation-state threat actor.

Another similar incident occurred on June 25, 2013, on the 63rd anniversary of the start of the Korean War (1950–1953), which resulted in the division of the Korean peninsula. On that day, multiple attacks reminiscent of nationalistic hacktivism, a type of patriotic activism, targeted the Blue House, government ministries, and media by defacing web pages, stealing data, and corrupting servers. One of the distributed denial-of-service (DDoS) attacks observed against the South Korean government websites was directly linked to malware used in the DarkSeoul attack [16]. The ontology in Section 4 does not account for a nationalistic hacktivist threat actor type that would ideally characterize this operation's actor. The defining attributes of each threat actor type describe their subtle differences. For example, even though the characterization attributes of the nationalistic hacktivist type would highly overlap with the radical activist type in terms of outcomes and objectives, nationalistic hacktivists are mainly motivated by the desire to achieve dominance over another nation because of their loyalty and strong devotion to their own nation or the leaders of the nation. In contrast, a radical activist operates for more ideological and political reasons to replace the fundamental principles of a society or a political system. In addition, nationalistic hacktivists would be resource-constrained compared to a nation-state-backed group. As explained in Section 3, the definition of new actor types and updating existing ones should be a standards-based task where the security community agrees on explainable characterization attribute-based descriptions for promoting and facilitating universal adoption.

In November 2014, Sony Pictures Entertainment (SPE) was attacked with malware resulting in information theft which was later used for extortion regarding canceling the release of a film depicting an assassination plot against North Korean leader Kim Jong Un. The stolen data included employee personal information, company emails, usernames and passwords, details of SPE's internal IT infrastructure, and unreleased movies. In addition, the attackers succeeded in rendering thousands of computers inoperable by deleting the master file table and the master boot record from hard drives [17]. The perpetrators identified themselves as Guardians of Peace (GOP). The attack, which was initially believed to be the work of a hacktivist group or disgruntled insiders, was later attributed to the Lazarus Group [18]. Based on available information, we characterize the operation and derive the following attributes. The Sony incident was a covert operation (*visibility* → *covert*) planned and executed by an unknown external group (*access* → *external*) that caused theft of information and damage to assets (*objective* → *copy, damage, destroy*). The stolen information was used to hurt the company's image and resulted in significant financial losses (*outcome → damage, embarrassment*). The extortion demands, in addition to threatening emails sent to Sony employees, reflected a threat actor who takes no account of the law (*limits → extra-legal, major*) and an actor who attempts to achieve dominance through its actions (*defining motivation → dominance*). In addition, the threat actor demonstrated considerable resources and advanced skills, as indicated by its persistence in Sony's network and the significant losses suffered (*skills → adept*), (*resources → at least organization*). Based on the above characterization, a reasoner would infer that the populated attributes are equivalent to government cyberwarrior or otherwise known as a nation-state threat actor type. Nevertheless, the attack could also be understood as a form of nationalistic hacktivism because of its context. Interestingly, in the early stage of the attack and before the explicit demand to withdraw the movie's theatrical release, some of the targeted high-ranking Sony employees received compensation requests from the attackers for the damage they had suffered [17]. This could indicate a personal financial motivation, irrespective of the group's primary goal.

The Lazarus Group, being polymorphic, has also been observed to be financially motivated and has demonstrated highly organized and sophisticated cyber criminal behavior by penetrating targets with large financial streams. According to Kaspersky [11], Lazarus Group operations are expensive, and financially motivated attacks could be a way to better finance them. Chanlett-Avery et al. emphasized that the Lazarus Group engages in financially motivated attacks to raise revenue for the regime in response to sanctions imposed by the United States and the United Nations Security Council as a reaction to North Korea's weapons of mass destruction and ballistic missile programs, as well as human rights abuse [19].

Temporality-based semantic representation and inference provide more complete, queryable, and explainable intelligence and a certain extent of automation in intelligence generation with respect to how threat actors evolve into new behaviors. Based on the queries that an organization wants to answer, the characterization attributes and inferred information (instances) can be used to derive highly relevant and contextual cyber threat intelligence. Furthermore, universally agreed unambiguous definitions and vocabularies enable more robust information sharing.

As illustrated by Figure 9, the evidence indicates that the Lazarus Group is polymorphic and, through its operations, has exhibited behavior and capability aligned with organized cyber crime, nationalistic hacktivists, cyber vandals, and nation-state-backed entities.

**FIGURE 9:** THE POLYMORPHISM OF THE LAZARUS GROUP

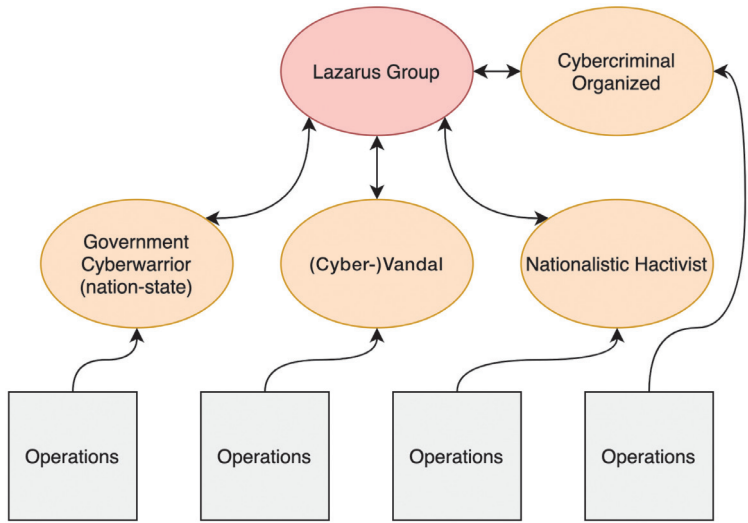

## 6. CONCLUSION

Threat actors are becoming increasingly sophisticated and polymorphic. To understand those hybridized threats, defenders seek timely, accurate, relevant, and actionable threat intelligence for anticipatory threat reduction. Today's threat intelligence tends to be ambiguous and inadequately structured to track and demystify changes in the behavior of actors over time, such as new goals, motivations, and related operations and TTPs. Threat actors have an asymmetric information advantage over defenders. Before executing a targeted attack, they are well aware of the profiles, infrastructures, systems, and applications of their victims. This work laid the foundation for generating highly contextual, explicable, processable, and shareable threat actor intelligence that can accurately capture, interpret, and explain changes in threat actor behavior and their polymorphism over time. In particular, we demonstrated how a set of characterization attributes can enrich threat actor information and how, in combination, can enumerate their type. By encapsulating this knowledge within an ontology, we demonstrated how a perpetrator's nature could be inferred automatically using deductive reasoning and withhold the relations/semantics that justify the inference.

# ACKNOWLEDGMENTS

The authors would like to express their gratitude to Mr. Paul Patrick from DarkLight Inc. for providing comments that helped improve the manuscript.